\documentclass[a4paper, reqno]{amsart}
\usepackage[latin1]{inputenc}
\usepackage{graphicx}
\usepackage{natbib}

\usepackage{amsmath}
\usepackage{amssymb}
\usepackage{amsthm}
\usepackage{amsaddr}

\numberwithin{equation}{section}
\calclayout

\begin{document}
\title[Persistence of lytic bacteriophages in vivo]{On the persistence of lytic bacteriophages in vivo and its consequences for bacteriophage therapy}
\author{Matthias M. Fischer}
\address{Freie Universit\"at Berlin, Institut f\"ur Biologie, Mikrobiologie, K\"onigin-Luise-Strasse 12-14, 14195 Berlin, Germany.}
\email{m.m.fischer@fu-berlin.de}
\date{\today}

\begin{abstract}
Bacteriophages are viruses infecting bacteria and archaea. Many phage species cause infections which lead to the certain death of the infected prokaryotic host cell and the release of a large batch of phage progeny, yet they have been able to stably coexist with their bacterial hosts over the eons in nature, as well as in the majority of laboratory experiments reported in the literature. This possibility of a stable coexistence between populations of bacteria and bacteriophages can be suspected to critically reduce the chances of a successful therapeutic application of bacteriophages for combatting pathogenic bacterial infections \textit{in vivo}. Here, we extend an established differential equation model describing the interaction of bacteria, bacteriophages, and the host immune system, modelling different degrees of spatial heterogeneity of the host organism by introducing a scaling parameter which alters the encounter rates of the different cell populations. We demonstrate by rigorous mathematical analysis that, depending on the degree of spatial heterogeneity, the system will either converge to the desired state of bacterial and phage extinction, to a state of persisting bacterial infection with a completely eliminated bacteriophage population or to an equally undesirable state of a stable long-term bacteria-bacteriophage coexistence. We additionally provide numerical solutions of the model to illustrate the emerging dynamics and discuss some implications for bacteriophage therapy.
\end{abstract}

\maketitle

\section{Introduction}
\label{intro}
Bacteriophages are viruses that infect and replicate within bacteria and archaea. They can either have only a lytic lifecycle or both a lytic and a lysogenic one. In case of a lytic life cycle, the genome of the bacteriophage is replicated and expressed as proteins within the bacterial host cell after a successful infection. After the assemblage of the virions, the host cell is lysed and the assembled particles are released. In contrast, in case of a lysogenic life cycle the virus persists as a prophage by integrating into the host genome or is maintained extrachromosomally in the host cell \citep{brockmadigan}. A third developmental route, called 'pseudolysogeny', has also been proposed, in which the lysis of the host cell is postponed over the course of one or more host generations without the formation of a prophage, which primarily seems to occur in cases of nutrient-depleted host cells \citep{pseudo2}. As pointed out by \citet{Heilmann2}, many phage species solely rely on a virulent lifecycle, causing any infection to lead to the certain death of the infected host cell and the release of a large batch of phage progeny. Consequently, it is a bewildering question how such voracious parasites have been able to stably coexist with their bacterial hosts over the eons. Adding to one's astonishment, in the great majority of experimental studies reported in the literature, a long-term persistence of the phage population in a mixed bacteria-bacteriophage culture could be observed (for example, refer to \citet{t4r_persistence}, \citet{t7_persistence} or \citet{tseries_persistence}), while cases of phage extinction are in fact rather rare \citep{persistence}. \\

Bacteriophages have been proposed as a possible way of treating pathogenic bacterial infections in humans or animals since the mid of the last century \citep{phages}, because they offer a number of advantages over classical antibiotic therapy: By virtue of being significantly more specific to their bacterial hosts than antibiotics, they can be expected to be generally harmless to the patient as well as the microbiomes of the patients, and to have a lower amount of undesirable side effects \citep{sideeffects1, sideeffects2}. Due to the increasing rates of bacterial resistances towards established antibiotics \citep{resistances1, resistances2}, as well as the steady decline in the numbers of newly approved antibiotics \citep{decline1, decline2}, bacteriophage therapy has again become a focus of attention in biomedical research -- for an example of a recent clinical trial refer to \citet{trial}. Nonetheless, the possibility of a stable coexistence between populations of bacteria and bacteriophages, as outlined previously, may in some cases very well be able to critically reduce the chances of a successful therapeutic application of bacteriophages \textit{in vivo}. Accordingly, understanding the exact mechanisms enabling such a coexistence can be expected to be of great importance for a successful clinical translation. \\

Generally, two major reasons for the possibility of such a long-term coexistence between host and parasite have been proposed: First, a coevolution between the two interacting populations in the form of an 'evolutionary arms-race', or, put more formally, a \textit{reciprocal genetic change in interacting species, owing to natural selection imposed by each on the other} \citep{Futuyma}. This has been studied e.g. by \citet {Weitz} using sets of differential equations extended by an event-driven system based on random numbers, where traits of the interacting organisms were varied, and for every newly arising genotype, a new equation was added to the system. They were able to show that multiple strains of bacteria and bacteriophages can permanently coexist in a homogeneous medium with a single resource and that diversification leads to different genotypes of phages that adsorb effectively to only a limited subset of all existent bacterial genotypes. In other words, functional differences between bacterial and viral genotypes will arise endogenously within the simulated cocultures. \\

The other proposed mechanism enabling such a stable coexistence refers to effects of the spatial environment on the interacting populations. In case of a typical continuous bacteria-bacteriophage coculture taking place in a well-mixed chemostat system, this is primarily possible due to the formation of a bacterial biofilm on the inside of the reaction vessel, in which so-called wall populations can persist next to the populations in the planktonic phase. This has for example been experimentally demonstrated by \citet{wall1} who showed that cocultures with bacterial wall populations allowed long-term coexistence in contrast to cocultures taking place in a completely homogeneous planktonic system. They also demonstrated that the \textit{final percentage of sensitive bacteria was significantly higher in walls than in liquid populations}, concluding that biofilms on glass surfaces may also act as spatial refuges for bacteria with a genotype not resistant to bacteriophage infection. In addition to acting as a refuge for sensitive bacteria, the complex spatial microstructure of biofilms can be assumed to directly influence the infection likelihood per host encounter and the diffusion ability of phages. This has, for example, been suggested by the experimental results of \citet{Brockhurst} who studied the interaction between the bacteriophage PP7 and the bacterium \textit{Pseudomonas aeruginosa} in spatially heterogeneous cultures. A study by \citet{Luckinbill} on the interaction of \textit{Paramecium aurelia} and its predator \textit{Didinium nasutum} has further substantiated this idea by demonstrating that the two populations were able to coexist if and only if methyl cellulose was added to the growth medium leading to slower dispersal of the organisms, thereby causing lower encounter rates of predator and prey species. The same has already been pointed out by \citet{Simmons} who derived an agent-based simulation framework of the bacteria-bacteriophage interaction in a growing biofilm. They determined the phage mobility in the biofilm phase as one of the core determinants for a long-term stable coexistence and demonstrated that the probability of a successful elimination of the bacterial population significantly decreases if phage mobility in the biofilm phase is reduced. Since the authors modelled the interaction in an \textit{in vitro} culture, these results might not be completely transferable to the medically relevant \textit{in vivo} case, where additional effects of the host immune systems are at play. Additionally, the results of such agent-based simulations should primarily be understood as a set of observations on a series of \textit{in silico} experiments, lacking the inherent strengths of traditional mathematical models which allow to definitely 'prove' the observed relationships \citep{edelstein}. \\

Hence, in this work we provide a rigorous mathematical analysis based on an established differential equation model by \citet{leung} describing the \textit{in-vivo} interaction of a bacterial and a bacteriophage population as well as the innate host immune system of the infected organism. We implicitly model different degrees of spatial heterogeneity and the complexity of the spatial microstructure of the system by introducing a scaling parameter which alters the encounter rates of the different cell populations. We calculate the occurring equilibrium states of the system and assess their asymptotic stability for varying values of this scaling parameter, and will demonstrate that the degree of spatial heterogeneity crucially affects the qualitative behaviour of the system: Depending on the degree of spatial heterogeneity, the system will either converge to the desired state of bacterial and phage extinction, to the state of a persisting bacterial infection with a completely eliminated bacteriophage population or to a state of a stable long-term bacteria-bacteriophage coexistence.

\section{Materials and methods}

\subsection{Mathematical model}
We base our analysis on an established model by \citet{leung}. Since we are mainly interested in the potential application of bacteriophages as a means of treating bacterial infections which cannot be dealt with by the immune system of the host alone, we can assume that the density of immune cells $I(t)$ has already grown to the maximum possible value of $K_I$, so that $I(t) = K_I = const.$ This reduces their three-dimensional system to a two-dimensional one, describing the rates of change of the bacteria and phage population densities as follows:

\begin{equation}
\begin{split}
\dot{B}(t) &= \frac{K_B - B(t)}{K_B} r B(t) - \sigma \Phi B(t) P(t) - \sigma \epsilon B(t) K_I / \left(1+\frac{B(t)}{K_D}\right)\\
\dot{P}(t) &= \beta \sigma \Phi B(t) P(t) - \omega P(t) \\
\end{split}
\end{equation}

where $K_B, K_I$ describe the maximum carrying capacity of the host for the bacterial and the immune cell population, respectively. The maximum growth rate of the bacterial population is termed $r$; the constant $K_D$ describes the bacterial cell concentration at which the immune response is half saturated to account for the immune evasion at higher bacterial concentrations. Following the original paper, the phage adsorption rate is labelled $\Phi$ and the killing rate of the immune system is termed $\epsilon$. The effective viral burst size is denoted with $\beta$, and $\omega$ denotes the phage clearance rate \textit{in vivo}. For the estimated values of the parameters refer to Table 3 in the original paper. We have also added a scaling parameter $\sigma$ as an additional factor to all encounter rates which we will manipulate later in order to simulate different levels of spatial heterogeneity. For $\sigma = 1$ the system is equivalent to the model presented in the original paper, while decreasing the parameter reduces the rates with which the different cell populations encounter each other.

\subsection{Numerical solutions}
Numerical solutions of the system were obtained using the Python programming language, version 3 \citep{python}, and the command {\tt solve\_ivp} in the Python software package {\tt SciPy} \citep{scipy}. All plots in this paper have been generated using the {\tt matplotlib} software library for Python \citep{matplotlib}.

\section{Results}
\subsection{Analytical examination}

We limit our search for steady states to the biologically feasible region, where the three variables are all equal or greater than zero. Three such equilibria can be found: First, the state of extinction $E$ with $\bar B = \bar P = 0$. Second, the phage-free equilibrium $F$ with 

\begin{equation}
\bar B = -\frac{K_C+K_D}{2} + \sqrt{\frac{(K_C-K_D)^2}{4} + K_C K_D \left( 1 - \frac{\sigma \epsilon K_I}{r} \right)}; \; \bar P = 0.
\end{equation}

From $\bar B$ one may conclude that meaningful solutions only arise for $\sigma \leq \frac{r}{4\epsilon K_I} \left( \frac{K_C}{K_D} + \frac{K_D}{K_C} + \frac{2}{r} \right) \approx 58.0$. Third, the coexistence state $C$ with 

\begin{equation}
\bar B = \frac{\omega}{\beta \sigma \Phi}; \; \bar P = \frac{r}{\sigma \Phi} \left(1-\frac{\omega}{\beta \sigma \Phi K_B} \right) - \frac{\epsilon K_I}{\frac{\omega}{\beta \sigma K_D}+\Phi},
\end{equation}

which importantly only yields feasible solutions for $\bar P$ if $\sigma \in [2 \times 10^{-4}; 0.587]$ under the standard parametrisation. Thus, a coexistence equilibrium starts to exist for a spatial scaling parameter of 0.587 or below. \\

The Jacobian of the system is given by

\begin{equation}
\label{eq:J2}
\textbf{J} = \left[ \begin{matrix}
-\frac{2r}{K}B(t) - \sigma \Phi P(t) -\frac{\epsilon \sigma K^2_D K_I}{(B(t)+K_D)^2} + r & - \sigma \Phi B(t) \\[1ex]
\beta \sigma \Phi P(t) & \beta \sigma \Phi B(t) - \omega \\[1ex]
\end{matrix}
\right],
\end{equation}

which at the extinction state $E$ has the eigenvalues $\lambda_1 = r - \epsilon \sigma K_I$ and $\lambda_2 = - \omega$. Accordingly, the extinction state is stable as long as the scaling parameter $\sigma > \frac{r}{\epsilon K_I} \approx 0.508$. This makes sense biologically as infections with small amounts of bacteria can normally easily be dealt with by the immune system, which makes the disease-free state biologically stable in praxis. Importantly, however, in case of increasing spatial complexity, e.g. due to the formation and growth of a biofilm or due to multiple compartments being involved in the infection, the extinction state loses its stability. \\ 

At the phage-free equilibrium $F$, we get the two eigenvalues 

\begin{equation}
\lambda_1 = -\frac{2r}{K} \bar B + r - \frac{\epsilon \sigma K^2_D K_I}{(\bar B + K_D)^2} ; \; \lambda_2 = \beta \sigma \Phi \bar B - \omega. 
\end{equation}

\begin{figure}[h!]
	\centering
	\includegraphics[width=12cm]{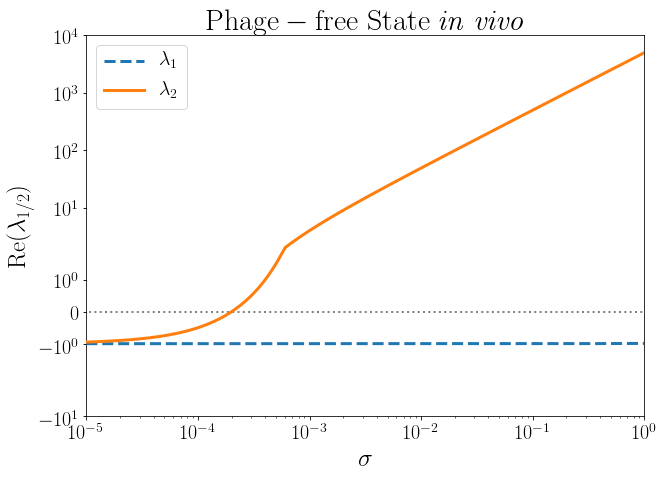}
	\caption{\textit{Real parts of the two eigenvalues of the Jacobian at the phage-free equilibrium for different values of the scaling factor $\sigma$. Observe how this state of eradication of the phage population becomes stable for a sufficiently small value of $\sigma$. Imaginary parts of zero not shown.}}
	\label{fig:invivo_F}
\end{figure}

One can easily show that under the standard parametrisation of the model, the first eigenvalue will stay below zero even for drastic changes in $\sigma$ spanning over several orders of magnitude. However, the second eigenvalue is crucially affected by the choice of $\sigma$. Figure \ref{fig:invivo_F} shows the real parts of the eigenvalues using the standard parametrisation introduced earlier as a function of the scaling parameter $\sigma$. Observe that given a sufficiently small scaling factor $\sigma$, the phage-free equilibrium will become stable, indicating a persistent bacterial infection which the host immune system cannot take care of. We do not show the imaginary part of the Jacobian here, as it is constantly equal to zero, which makes the equilibrium state a stable node or a saddle point respectively. \\

\begin{figure}[h!]
	\centering
	\includegraphics[width=12cm]{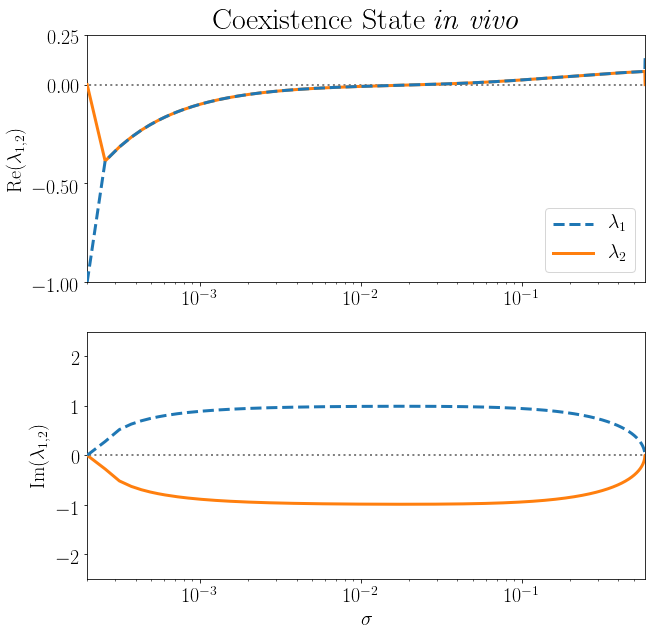}
	\caption{\textit{Real and imaginary parts of the two eigenvalues of the Jacobian at the coexistence equilibrium for different values of the scaling factor $\sigma$. The range of the x-axis is confined to the region in which the coexistence state C exists and has positive solutions.}}
	\label{fig:invivo_C}
\end{figure}

Finally, at the coexistence state $C$ we get complicated expressions for the eigenvalues which we omit here for brevity purposes. Figure \ref{fig:invivo_C} shows the real and imaginary parts of the two eigenvalues of the Jacobian of the system at the coexistence state. Note the existence of a range of values for $\sigma$, in which the real parts of both eigenvalues fall below zero, indicating the local stability of the state of bacteria-bacteriophage coexistence in the form of a stable spiral. For bigger values of the parameter, this spiral will become unstable until it finally vanishes at $\sigma \geq 0.587$. \\

\subsection{Representative phase portraits}
\begin{figure}[h!]
	\centering
	\includegraphics[width=15cm]{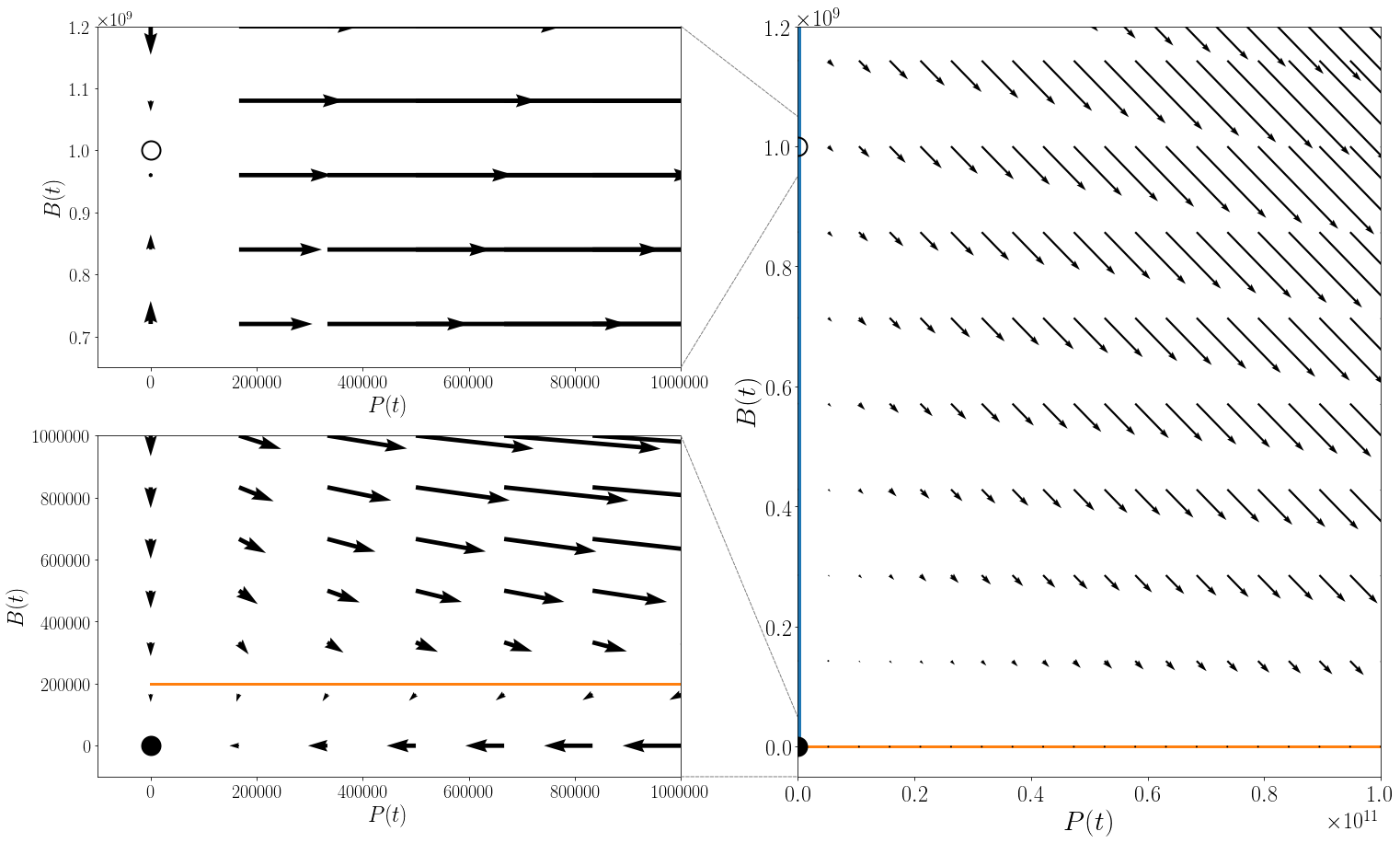}
	\caption{\textit{Phase portrait of the system for $\sigma = 1.0$. Black discs indicate stable equilibria, black circles unstable ones. Vertical blue line: non-trivial B-nullcline, horizontal orange line: non-trivial P-nullcline. Note that the non-trivial B-nullcline slightly falls into the second quadrant of the plot, which is the reason why it cannot be seen in the bottom left-hand side plot.}}
	\label{fig:sigma1}
\end{figure}

We now show representative phase portraits for different values of the spatial scaling parameter $\sigma$ in order to illustrate the resulting system. Figure \ref{fig:sigma1} shows the resulting phase space for a scaling parameter of $\sigma=1$, i.e. for the standard parametrisation. The phage-free state of bacterial infection is asymptotically unstable towards perturbations in the phage density (top-left plot, black circle), however the extinction state is stable (bottom-left plot, black disc). The state of coexistence does not yet exist due to $\sigma$ being greater than 0.587, causing the $B$-nullcline (vertical blue line) to fall into the second quadrant. Accordingly, a bacterial infection can be treated by introducing a sufficient amount of bacteriophages into the system. \\

\begin{figure}[h!]
	\centering
	\includegraphics[width=15cm]{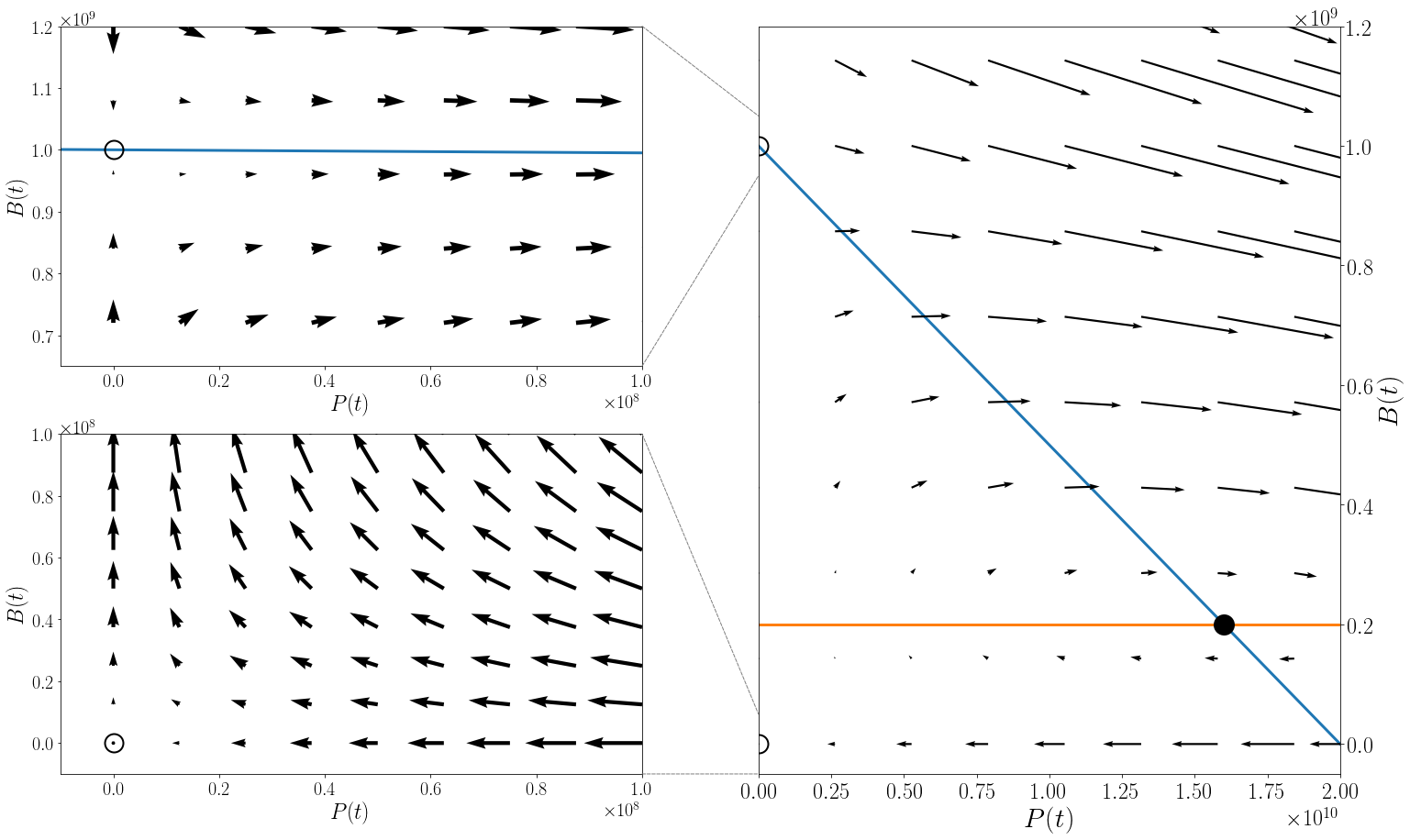}
	\caption{\textit{Phase portrait of the system for $\sigma = 10^{-3}$. Black discs indicate stable equilibria, black circles unstable ones. Diagonal blue line: non-trivial B-nullcline, horizontal orange line: non-trivial P-nullcline.}}
	\label{fig:sigma1e-3}
\end{figure}

Figure \ref{fig:sigma1e-3} shows the resulting phase space for a scaling parameter of $\sigma=10^{-3}$. The phage-free state of bacterial infection is still asymptotically unstable towards perturbations in the phage density (top-left plot, black circle), however the extinction state now has lost its stability due to $\sigma$ being less than 0.508 (bottom-left plot, black circle). A new state of coexistence in the form of a stable spiral has now appeared (black disc on the right-hand side at the intersection of the two nullclines), since $\sigma$ now falls below the critical value of 0.508. Accordingly, the introduction of bacteriophages will now shift the system from a state of bacterial infection to a state of stable bacteria-bacteriophage coexistence. \\

\begin{figure}[h!]
	\centering
	\includegraphics[width=15cm]{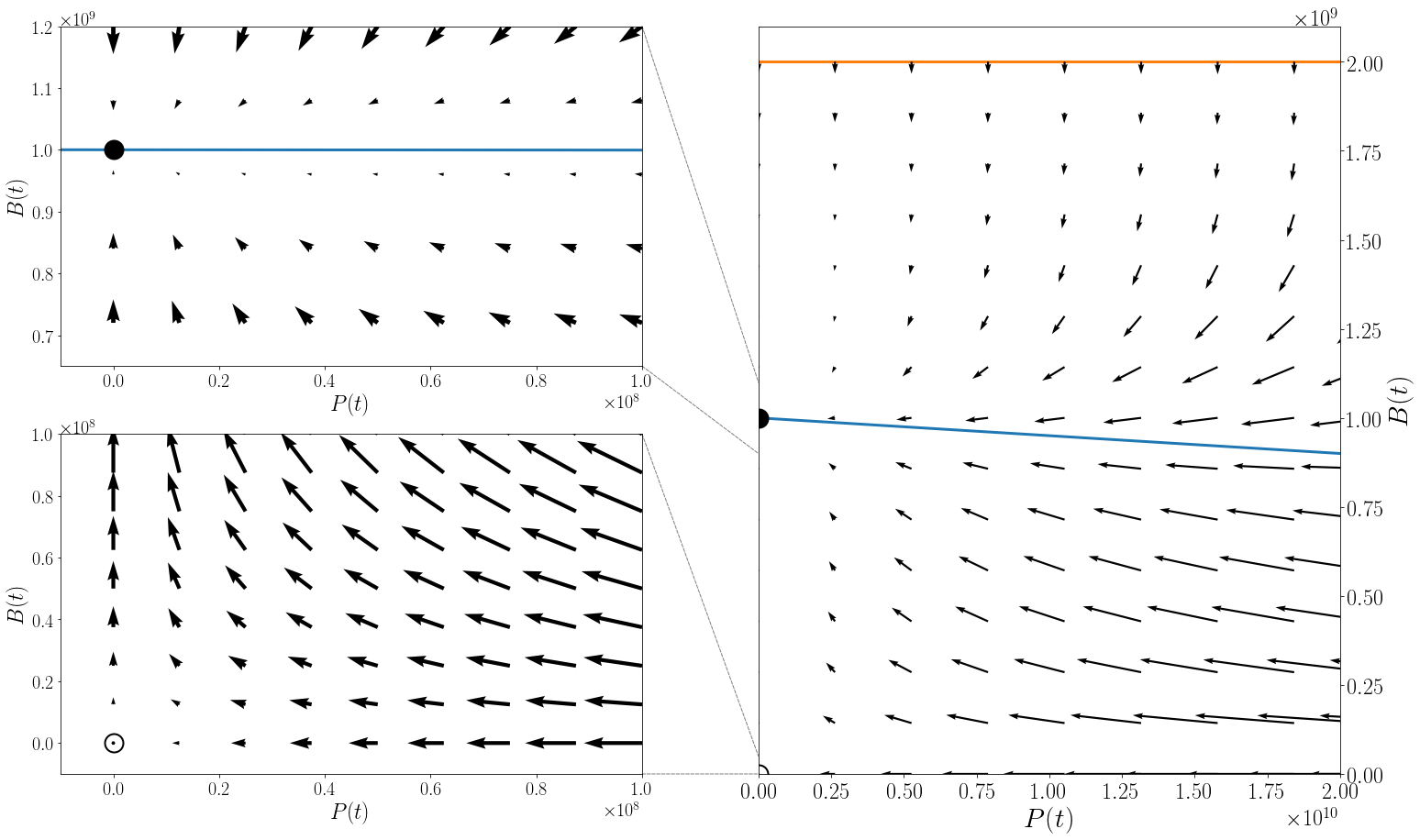}
	\caption{\textit{Phase portrait of the system for $\sigma = 10^{-4}$. Black discs indicate stable equilibria, black circles unstable ones. Diagonal blue line: non-trivial B-nullcline, horizontal orange line: non-trivial P-nullcline.}}
	\label{fig:sigma1e-4}
\end{figure}

Finally, figure \ref{fig:sigma1e-4} depicts the phase space for $\sigma=10^{-4}$. The state of phage-free bacterial infection is now stable (top-left plot, black disc), while the extinction state is unstable just as before (bottom-left plot, black circle). The state of stable coexistence now has vanished from the biologically meaningful first quadrant, since $\sigma$ has fallen below the critical value of $2 \times 10^{-4}$. Accordingly, the bacterial infection now is persistent and cannot be treated by bacteriophage therapy anymore. \\

\subsection{Numerical solutions of the model}
\begin{figure}[h!]
	\centering
	\includegraphics[width=15cm]{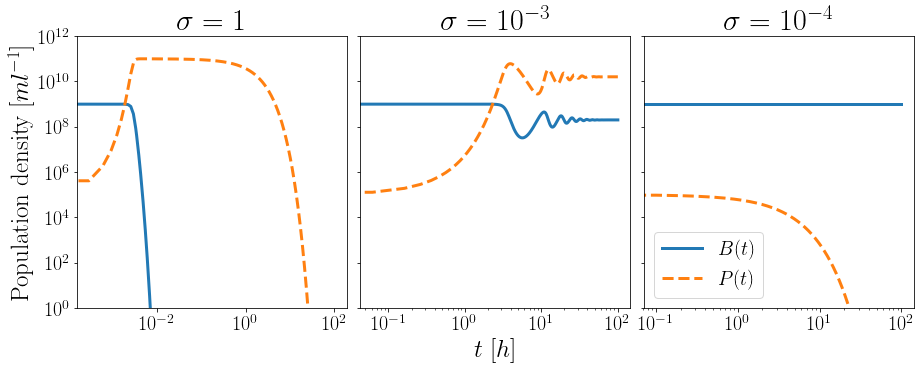}
	\caption{\textit{Numerical solutions of the system for different values of the spatial scaling parameter $\sigma$. The numerical approximations confirm what an inspection of the phase portraits has already suggested: With increasing spatial complexity, bacteriophages will not be able to eliminate the bacterial infection anymore, but will instead coexist stably with the bacteria (middle plot) or will be eliminated (right plot).}}
	\label{fig:sol}
\end{figure}

To illustrate the emerging dynamics of the system, we will now provide a number of numerical simulations of the derived model. In contrast to the simulations provided in \citet{leung}, we here assume the medically more meaningful scenario, in which the bacterial population density inside the host organism, as well as the immune response of the host have already reached the maximum value. We now add an inoculum of $10^5$ bacteriophages per millilitre into the host organism in order to try to combat the bacterial infection. Figure \ref{fig:sol} depicts the emerging dynamics for the same values of the spatial scaling parameter $\sigma$ as used in the previous section. The simulations confirm what an inspection of the phase portraits has already suggested: Depending on the degree of spatial complexity, the system will converge to one of the three equilibria respectively. It is again especially noteworthy that for a sufficiently high degree of complexity an eradication of the bacterial infection is not possible and instead either a stable coexistence of bacterium and phage will take place, or the bacteriophage population is eliminated from the system, while the bacterial infection persists. \\

\section{Discussion}
\label{sec:disc}
In this work, we have extended a simple and elegant ODE model by \citet{leung} to assess the influence of the spatial microstructure of a host organism on the \textit{in vivo} interaction of bacteria, bacteriophages and the host immune system. The degree of complexity of the spatial microstructure was modelled by introducing a scaling parameter which linearly scales the encounter rates between the three populations, making use of observations by \citet{Luckinbill}. This approach is especially appropriate in the light of the ability of bacteria to form biofilms -- hydrated matrices of extracellular polymeric substances with low diffusivity \citep{biofilm}, however the scaling parameter can also simply be thought of a parameter describing the general degree of spatial heterogeneity of the system. We have demonstrated changes in the qualitative behaviour of the system depending on this degree of complexity, namely the emergence of a stable equilibrium of long-term bacteria-bacteriophage coexistence, as well as the possibility of a stable bacterial infection not treatable by bacteriophage therapy. \\

Our results point into the same direction as the experimental work of \citet{Brockhurst} and analytically confirm and extend the results of the extensive agent-based simulations of the \textit{in-vitro} case by \citet{Simmons}, however offer the advantage of a simple, analytically examinable mechanistic mathematical model. It is especially noteworthy that the model presented in this work does not include any evolutionary processes such as the emergence of new genotypes. Accordingly, the stable states of coexistence does not seem to necessarily require any coevolutionary dynamics between the interacting populations, which, however, seem to be necessary for long-term coexistence in completely homogeneous environments as demonstrated by \citet{Weitz}. This finding is in agreement with the results of a recent work by \citet{gut} who found that the natural coexistence of bacteria and phages in the mouse gut was \textit{neither dependent on an arms race between bacteria and phages, nor on the ability of phages to extend host range.} Instead, they identified the spatial heterogeneity of the organ to be the core determinant of long-term coexistence. \\

The results of this study may also be understood as the examination of a mechanism that is possibly able to reduce the efficacy of bacteriophage therapy for the treatment of bacterial infections. Due to the high degree of spatial heterogeneity of eukaryotic organisms as well as the ability of bacteria to form biofilms, bacteriophage therapy might, in some cases, not eradicate the bacterial infection as desired, or lead to an undesirable stable long-term coexistence of bacteria and a bacteriophage populations in the patient. This can be expected to become especially relevant in case of complicated systemic infections, where multiple compartments of the host organism are affected. The fact that the exposure of bacteria to phages is also able to stimulate bacterial biofilm production as reviewed by \citet{biofilm_stim} may further complicate the matter, as well as interactions between the host immune system and bacteriophages which might increase the degradation rates of the latter, thereby decreasing their efficacy for controlling the bacterial infection of the host organism \citep{immune1, immune2}. \\

An interesting property of some bacteriophages is their ability to break down biofilms by carrying surface enzymes which can degrade bacterial polysyccharids \citep{degrade}, which appears to be a promising starting point to deal with the problem of biofilm-induced spatial heterogeneity in bacteriophage therapy. (For a more extensive review on the topic of bacteriophage-biofilm interaction, refer to \citet{phage_bf_interaction}.) As one might expect, the concentration of phage inoculum crucially influences the degree of biofilm disruption observable \textit{in vitro}, however even a high multiplicity of infection of 100 does not seem to guarantee bacterial extinction, but can still lead to a stable coexistence of phage and bacterium instead \citep{degrade_moi}. Additionally, viral surface depolymerases have been found to be rather specific and usually only able to target only a few, closely related bacterial polysaccharide structures \citep{specifity1, specifity2}. Finally, biofilms which consist of multiple bacterial species, can be expected to be more resistant to bacteriophage exposure than single species ones, as demonstrated in a study by \citet{multispecies}, which might be caused by additional structural heterogeneity within the biofilm. Experimental studies attempting to use phages to control bacterial biofilms \textit{in vivo} do not seem to have unanimously yielded promising results yet \citep{bf_invivo_1, bf_invivo_2}, and one may reasonably assume that in some cases other means of controlling chronic bacterial infections which have produced biofilms \textit{in vivo} may be required. Nonetheless, for some applications such as the treatment of locally confined chronic infections which are easily accessible from the outside, such as chronic diabetic ulcers, bacteriophage therapy indeed seems to be a promising approach \citep{diabetic}.

\pagebreak 

\section{Display items}

\clearpage

\section{Acknowledgements}
I would like to thank Matthias Bild and Prof. Dr. R. Mutzel for their guidance through the field of experimental and theoretical bacteriophage research during my years as an undergraduate student at Freie Universit\"at Berlin. I am also indebted to M.Sc. Joscha Reichert for valuable feedback regarding the preparation of the display items, and Dr. C. v. Toerne for fruitful mathematical discussions and helpful feedback to an earlier version of this manuscript.

\bibliographystyle{apalike}
\bibliography{references}
\clearpage
\pagebreak

\end{document}